\documentclass[conference, a4paper]{IEEEtran}
\usepackage{cite}
\usepackage{amsmath,amssymb,amsfonts}
\usepackage{bm}
\usepackage{algorithmic}
\usepackage{algorithm}
\usepackage{graphicx}
\usepackage{textcomp}
\usepackage{xcolor}
\usepackage{subfigure}

\def\BibTeX{{\rm B\kern-.05em{\sc i\kern-.025em b}\kern-.08em
    T\kern-.1667em\lower.7ex\hbox{E}\kern-.125emX}}
\begin{document}
\title{Physical Layer Security for NOMA-Enabled Multi-Access Edge Computing Wireless
	Networks\\}
\author{\IEEEauthorblockN{Yating Wen$ ^{\dagger \ddagger}$, Tong-Xing Zheng$ ^{\dagger \ddagger}$, Yongxia Tong$ ^{ *}$,  Hao-Wen Liu$ ^{\dagger \ddagger}$, Xin Chen$ ^{\dagger \ddagger}$, Pengcheng Mu$ ^\dagger $, Hui-Ming Wang$ ^\dagger $}
	\IEEEauthorblockA{$ ^\dagger $School of Information and Communications Engineering, Xi’an Jiaotong University, Xi’an 710049, P. R. China\\		
		$ ^\ddagger $National Mobile Communications Research Laboratory, Southeast University, Nanjing 210096, P. R. China \\
	$ ^* $WuHan Maritime Communication Research Institute, WuHan 430070, P. R. China}}
\maketitle
\begin{abstract}
Multi-access edge computing (MEC) has been regarded as a promising technique 
for enhancing computation capabilities for wireless networks. In this paper, we 
study physical layer security in an MEC system where multiple users offload 
partial of their computation tasks to a base station simultaneously based on 
non-orthogonal multiple access (NOMA), in the presence of a malicious 
eavesdropper. Secrecy outage probability is adopted to measure the security 
performance of the computation offloading against eavesdropping attacks. We aim 
to minimize the sum energy consumption of all the users, subject to constraints 
in terms of the secrecy offloading rate, the secrecy outage probability, and 
the decoding order of NOMA. Although the original optimization problem is 
non-convex and challenging to solve, we put forward an efficient algorithm 
based on sequential convex approximation and penalty dual decomposition. 
Numerical results are eventually provided to validate the convergence of the 
proposed algorithm and its superior energy efficiency with secrecy 
requirements.   
\end{abstract}
\section{Introduction}
With the rapid development of wireless 
networks, recent years have witnessed an unprecedented proliferation of smart wireless devices and ultra-low-latency applications. Nevertheless, a large number of smart wireless devices have limited 
computation capabilities such that they can hardly support those computation-intensive and latency-sensitive applications. 
Multi-access edge computing (MEC) has emerged as an appealing solution to overcome the above problem \cite{ref1}. By deploying MEC servers at the edge of wireless networks, e.g. base stations (BSs), wireless devices can offload their computation-heavy tasks to the BSs for remote execution. However, due to the broadcast nature of wireless communications, the computation offloading process is vulnerable to eavesdropping attacks. Therefore, it is crucial to take the security issue into account when designing an MEC wireless network.

Physical layer security (PLS), as a novel low-complexity security mechanism for safeguarding information security at the physical layer \cite{ref2}, \cite{ref3}, has recently been exploited to thwart eavesdropping attacks for MEC networks. For example, the authors in \cite{ref4} first proposed to employ the PLS to secure the wireless computation offloading for a multi-user and multi-carrier MEC network. The authors therein minimized the weighted sum energy consumption via jointly designing the optimal transmit power and multi-carrier allocation while subject to a secrecy constraint that the offloading rate of each user should not exceed its secrecy rate. The secure computation offloading was later examined in \cite{ref5} for an unmanned-aerial-vehicle (UAV) MEC network in the presence of both active and passive eavesdroppers.

More recently, PLS has also been investigated in non-orthogonal multiple access (NOMA)-enabled MEC networks, where NOMA is employed to improve the computation and energy efficiencies for the MEC network by stimulating multiple users communicate with the BS at the same time \cite{ref6}--\cite{ref8}. Specifically, considering a two-user MEC network, the authors in \cite{ref6} studied the problem of minimizing the weighted sum energy consumption subject to a secrecy outage probability constraint, and the authors in \cite{ref7} further examined the delay minimization problem. When it comes to a multi-user MEC network, the analysis and optimization become much more complicated. A very recent work \cite{ref8} has investigated the computation efficiency maximization problem with multiple users for a NOMA-MEC network. However, the results obtained in \cite{ref8} are mainly based on two ideal assumptions. First, the eavesdropper's channel state information (CSI) is assumed to be perfectly known, which is not practical since the eavesdropper is usually a passive listener. In addition, instead of designing the optimal decoding order of successive interference cancellation (SIC) for the NOMA scheme, the authors therein have simply adopted the descending order according to the channel gains of the users.

Motivated by the aforementioned endeavors, in this paper, we investigate PLS for a multi-user NOMA-MEC network without eavesdropper's instantaneous CSI. Moreover, we study the secure resource allocation problem for minimizing the sum energy consumption by jointly optimizing the SIC decoding order, the local computing bits, the transmit power, and the rates for both the codeword and the confidential data. In order to address the above sophisticated non-convex problem, we develop an effective algorithm based on sequential convex approximation (SCA) and penalty dual decomposition (PDD).

\section{System Model and Problem Formulation}
\begin{figure}[!t]
	\centering
	\includegraphics[scale=0.4]{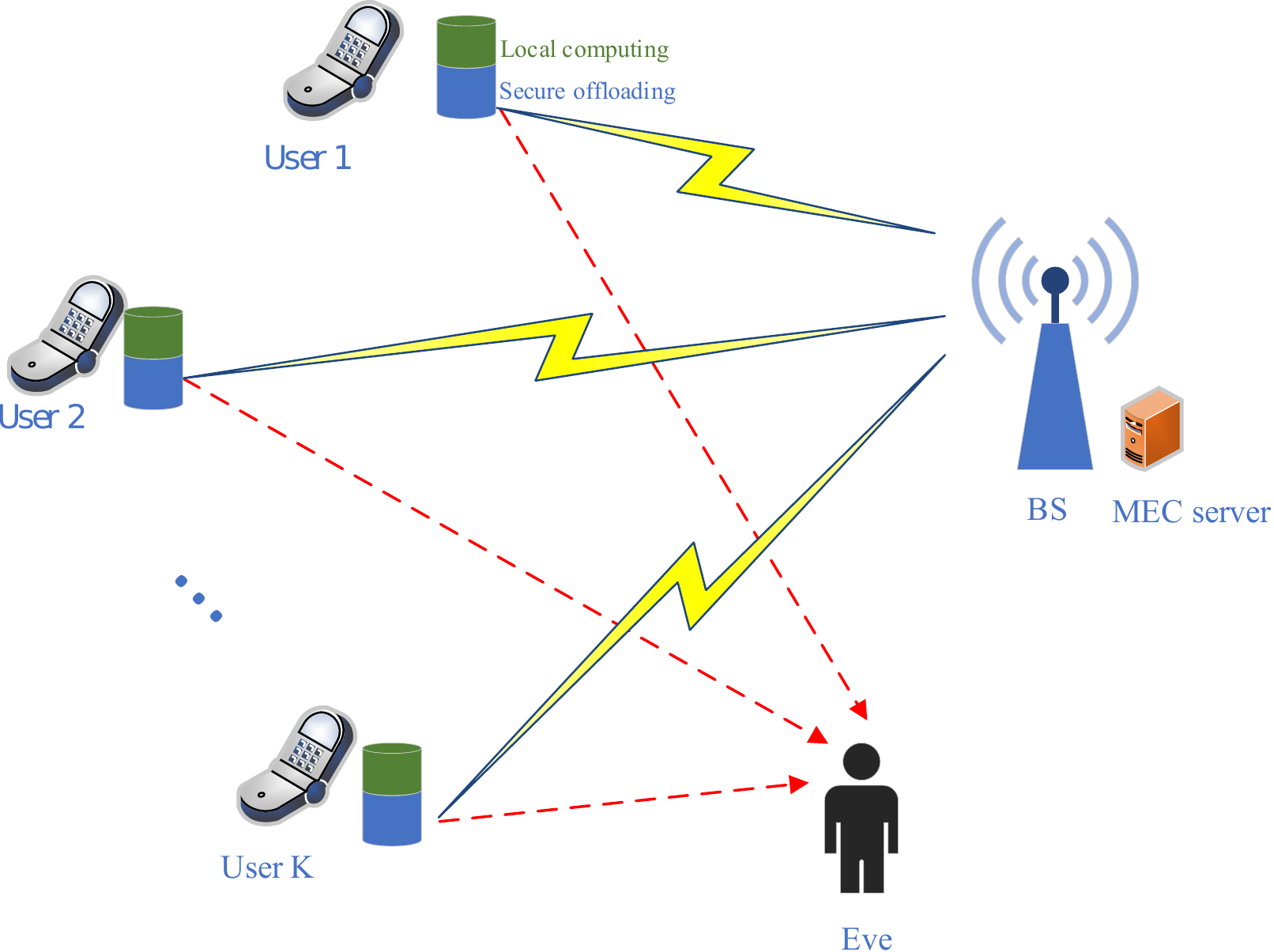}
	\caption{Illustration of a NOMA-MEC network with secure computation offloading from $ K $ users to a BS in the presence of an eavesdropper (Eve).}	
	\label{system model}
\end{figure}
As depicted in Fig. 1, we consider an uplink NOMA-MEC network consisting of a 
BS (with an MEC server integrated), $ K>1 $ users, and an external eavesdropper 
(Eve). All the nodes each are equipped with a single antenna. The users aim to 
securely offload their computation tasks to the BS while without being 
intercepted by Eve. We consider wireless channels which are subjected to 
frequency non-selective quasi-static block fading, and the channel 
coefficients from user $ k $ to the BS and to Eve are denoted by $ 
h_k=d^{-\alpha/2}_kg_k $ and $ h_{e,k}=d^{-\alpha/2}_{e,k}g_{e,k} $, 
respectively, for $ k\in \mathcal{K} \triangleq \left\{1, \cdots ,K\right\} $, 
where $ d_k $ and $ d_{e,k} $ denote the distances from user $ k $ to the BS 
and 
Eve; $ \alpha $ denotes the path-loss exponent; $ g_k $, $ g_{e,k} \sim 
\mathcal{C} \mathcal{N}(0,1) $ are the normalized Rayleigh fading channel 
states. Without loss of generality, $ K $ users are sorted in a descending 
order as per their channel gains such that 
$\left|h_1\right|^2>\left|h_2\right|^2>\cdots>\left|h_K\right|^2 $. We assume 
that the BS knows the instantaneous channel gains of all the users, i.e., $  
\left|h_k\right|^2 $, but can only acquire the average channel gain of Eve, 
i.e., $\mathbb{E} \left\{\left|h_{e,k}\right|^2 \right\} $\cite{ref9}. We also 
suppose that user $ k $ should finish its computation task of $ L_k $ input 
bits within a finite duration $ T $. Moreover, a partial offloading policy is 
considered, where each user processes a fraction of its computation task 
locally and offloads the remaining to the BS.

\subsection{Local Computing}
For user $ k\in \mathcal{K} $, partial of its computation task, i.e., $ l_k <L_k $ bits, are accomplished locally such that the rest $ L_k-l_k $ bits are offloaded to the BS. Let $ C_k $ denote the number of CPU cycles required for computing one 
input bit at user $ k $. Hence, the total number of CPU cycles required for the local computing at user $ k $ is $ C_kl_k $. For the sake of energy efficiency, for each CPU cycle $ n\in\left\{ 1,\cdots ,{{C}_{k}}{{l}_{k}} \right\} $, user $ 
k $ can control the CPU frequency $ f_{k,n} $ by employing the dynamic voltage and frequency scaling (DVFS) technique. As a consequence, the total execution time for the local computing of user $ k $ is 
$ \sum\nolimits_{n=1}^{{{C}_{k}}{{l}_{k}}}{\frac{1}{{{f}_{k,n}}}}$. In order to execute local computing within the duration $ T $, the CPU frequencies at user $ k $ should be chosen as $ f_{k,1}=\cdots 
=f_{k,{{{C}_{k}}{{l}_{k}}}} = {{{C}_{k}}{{l}_{k}}}/T$. Therefore, the energy 
consumption of user $ k $ for local computing is given by
\begin{equation}
\label{local}
E_k^{loc}=\sum\limits_{n=1}^{{{C}_{k}}{{l}_{k}}}{{{\varsigma 
}_{k}}{f_{k,n}^{2}}}=\frac{{{\varsigma }_{k}}{C_k^3}{l_k^3}}{T^2},\text{  
}\forall k\in \mathcal{K},
\end{equation}
where $ \varsigma_k>0 $ denotes the effective capacitance coefficient which depends on the chip architecture at user $ k $.

\subsection{NOMA-based Computation Offloading}
In our considered system, NOMA is employed to enable all the $ K $ users to offload their computation tasks to the BS 
using the same time and frequency resources. Under the NOMA offloading policy, the received signals at the BS and Eve
are expressed as 
\begin{align}
	y_{b}&={\sum\limits_{k=1}^K{\sqrt{p_k}}{h_k}{s_k}+{n_{b}}},\\
	y_e&=\sum\limits_{k=1}^K{{\sqrt{p_k}}{h_{e,k}}{s_k}+n_e},
\end{align}
respectively, where $s_k $ denotes the task-bearing signal of user $ k $ with $\mathbb{E}\left[ {{\left| {{s}_{k}} \right|}^{2}} \right]=1$, $ p_k>0 $ denotes the associated transmit power, and $ n_{b} $ and $ n_e $ denote the additive white Gaussian noise (AWGN) with zero mean and variances $ \sigma_{b}^2 $ and $ \sigma_e^2 $ at the BS and Eve, respectively.

Based on the NOMA policy, the BS will employ SIC to decode the signals received from the $ K $ users in a certain order. Specifically, the BS first decodes the signals with higher power while treating those weaker signals as interference. For convenience, we introduce a binary indicator ${\beta }_{k,l}\in \left\{0,1\right\} $ to describe the decoding order at the BS \cite{ref10}. To be specific, when the product of the relative channel gain $ {{\tau }_{k}} \triangleq \left|h_k\right|^2/\sigma_{b}^2 $ and transmit power $ p_k $ of user $ k $ is larger than that of user $ l $, we set $ \beta_{k,l}=1 $, which means that the signal of user $ k $ is decoded before that of user $ l $. $ {\beta }_{k,l}=0 $ indicates the opposite case. Thus, the binary indicator ${\beta }_{k,l} $ 
is given by
\begin{align}
	{\beta }_{k,l} &=\begin{cases}
		1,& {\tau_{k}}{p_k}>{\tau_{l}}{p_l}\\
		0,& {\tau_{k}}{p_k}<{\tau_{l}}{p_l},\quad \forall k,l \in \mathcal{K}\\
		0\  \mathrm{or} \ 1, & {\tau_{k}}{p_k}={\tau_{l}}{p_l}
	\end{cases}\\
	{{\beta }_{k,l}}&+{{\beta }_{l,k}}=1.
\end{align}

With the aid of the above binary indicator, the signal-to-interference-plus-noise ratio (SINR) of user $ k $ 
can be expressed as
\begin{equation}
	\gamma_k =\frac{{\tau_{k}}{p_k}}{\sum\nolimits_{l\in K,l\ne 
			k}{{\beta_{k,l}}{\tau_l}{p_l}+1}}, \forall k\in \mathcal{K}.
\end{equation}

In addition, the energy consumption of user $ k $ for computation offloading is given by
\begin{equation}
	E_k^{off}={p_k}{T}, \forall k\in \mathcal{K}.
\end{equation}
 
\subsection{Secure Encoding}
This subsection discusses the security issue of the computation offloading 
against eavesdropping. From a robust design perspective, we consider a 
worst-case scenario that Eve has a powerful multi-user decoding capability such 
that she can completely resolve the interference before decoding the desired 
signal. Therefore, Eve's received SINR for decoding the signal from user $ k $ 
can be given by
\begin{equation}
	\gamma_{e,k}=\tau_{e,k}{p_k},\forall k\in \mathcal{K} 
\end{equation}
where ${{\tau }_{e,k}}\triangleq{{\left| {{h}_{e,k}} \right|}^{2}}/{\sigma _e^{2}}$. 

We adopt the well-known Wyner's secrecy encoding scheme \cite{ref11} to secure computation offloading, where redundant information is intentionally added to confuse Eve. We denote the codeword rate and the confidential data rate for user $ k $ as $ R_{t,k} $ and $ R_{s,k} $, and the corresponding redundant information rate can be calculated as $ R_{e,k}=R_{t,k}-R_{s,k} $. We consider a practical case where the instantaneous CSI of Eve is unknown, and the secrecy outage probability is introduced to measure the secrecy performance of computation offloading. Mathematically,  the secrecy outage probability of user $ k $ can be defined as
\begin{equation}
\mathcal{P}_{so,k} \triangleq \mathrm{Pr}\left\{{R_{e,k} \le C_{e,k}}\right\},\forall k\in \mathcal{K}
\end{equation}
where $  C_{e,k} =\mathrm{log}_2(1+\gamma_{e,k}) $ denotes the maximum achievable rate of Eve for decoding the signal from user $ k $. The above definition means that only when $ R_{e,k} $ exceeds $ C_{e,k} $, the offloaded computation bits will not be decoded by Eve. Otherwise, the secrecy will be compromised, and a secrecy outage event is deemed to occur.

\subsection{Problem Formulation}
In this paper, we seek to design an energy efficient NOMA-MEC system by 
focusing on minimizing the sum energy consumption for all the users while guaranteeing successful and secure
computation offloading within a limited duration. The overall optimization problem can be formulated as follows
\begin{subequations}
	\label{optimal problem}
	\begin{align}
	\min_{\boldsymbol{l},\boldsymbol{p},\boldsymbol{R_t},\boldsymbol{R_s}, 
	\bm{\beta}} &\sum\limits_{k}{{{\varsigma 
			}_{k}}{{c}_{k}}^{3}l_{k}^{3}/{{T}^{2}}+{{p}_{k}}T},\\
	\mathrm{s.t.} \quad
	\label{rate}
	&BTR_{s,k} \ge L_k-l_k,\\
	\label{chca}
	&R_{t,k} \le C_{b,k},\\	
	&R_{t,k} \ge R_{s,k},\\	
	\label{secrecy}
	& \mathcal{P}_{so,k} \le \epsilon, \\	
	& p_k \ge 0,\\
	& 0\le l_k \le L_k,\\	
	\label{order}
	& {\beta }_{k,l} =\begin{cases}		
    1,& {\tau_{k}}{p_k}>{\tau_{l}}{p_l}\\
    0,& {\tau_{k}}{p_k}<{\tau_{l}}{p_l},\\
    0\ \mathrm{or}\ 1, & {\tau_{k}}{p_k}={\tau_{l}}{p_l}\\
    \end{cases}\\
   &{{\beta }_{k,l}}+{{\beta }_{l,k}}=1,
\end{align}
\end{subequations}
where $\boldsymbol{l}=[{l}_{1},\cdots,{{l}_{K}}]$ denotes the vector of local 
computation bits, $\boldsymbol{p}=[{p}_{1},\cdots,{{p}_{K}}]$ denotes the 
transmit power vector, $ \boldsymbol{R_t}=[R_{t,1},\cdots,R_{t,K}] $ denotes 
the codeword rate vector, $ \boldsymbol{R_s}=[R_{s,1},\cdots, R_{s,K}] $ 
denotes the confidential data rate vector, $ \bm{\beta} $ denotes the decoding 
order matrix, $ B $ denotes the system bandwidth, and $ C_{b,k}= 
\mathrm{log}_2(1+ \gamma_k) $ denotes the maximum achievable rate of the BS for 
decoding the signal from user $ k $. Constraint (\ref{rate}) implies that the 
confidential data rate should not be less than the offloading rate, such that 
the computation bits can be successfully offloaded to the BS within the 
duration $ T $ with bandwidth $ B $, constraint (\ref{chca}) ensures that the 
signal from user $ k $ can be decoded by the BS, constraint (10d) guarantees a 
positive redundant information rate against eavesdropping attacks, constraint 
(\ref{secrecy}) stands for a secrecy requirement where $ \epsilon \in 
\left(0,1\right) $ denotes the maximum tolerable secrecy outage probability, 
constraint (\ref{order}) ensures the stronger signal to be decoded first, and 
when two signals are equally strong, either one can be decoded first, and
constraint (10i) avoids the case where the signals from two users are decoded 
simultaneously at the BS.

Note that constraints (\ref{order}) and (10i) contain the binary variables $ \beta_{k,l} $ and constraint (\ref{secrecy}) is highly coupled with constraints (10b) and (\ref{chca}), and therefore problem (\ref{optimal problem}) is non-convex. In the next section, we will solve the above problem by proposing an efficient algorithm based on the SCA and PDD methods.
\section{Sum energy consumption minimization} 
In this section, we first transform the original problem (\ref{optimal problem}) into a more tractable form and then develop the solution by resorting to the SCA and PDD techniques.
\subsection{Problem Transformation}
Since the expression of SINR $ \gamma_k $ in (6) is in a fractional form, it is difficult to handle constraint (\ref{chca}). Therefore, we introduce two auxiliary variables $ b_k $ and $ \pi_k $ as 
the lower bound for $ \gamma_k $ and the upper bound for $ {\sum\nolimits_{l\ne k}{{\beta_{k,l}}{\tau_l}{p_l}+1}} $, respectively. By doing so, constraint (\ref{chca}) can be transformed as follows
\begin{subequations}
\label{chcatran}	
\begin{align}	
	R_{t,k}&\le 1+b_k,\\
	1+\sum\nolimits_{l\ne k}{{{\beta }_{k,l}}{{\tau }_{l}}{{p}_{l}}}&\le {{\pi 
	}_{k}},\\
	{b_k}{\pi_k}&\le {\tau_k}{p_k}.
\end{align}
\end{subequations}

In order to minimize the sum energy consumption, it is not difficult to determine that constraint (10c) should be active. Hence, by replacing $ R_{t,k} $ with $ C_{b,k} $, $ \mathcal{P}_{so,k} $ in (10e) can be calculated as
	\begin{equation}
	\begin{split}
	\mathcal{P}_{so,k}&=\mathrm{Pr}\left\{\mathrm{log}_2(1+\gamma_k)-R_{s,k}\le 
	\mathrm{log}_2(1+\gamma_{e,k})\right\}\\
	&=\mathrm{Pr}\left\{\left| h_{e,k} \right|^2\ge \theta_k \right\}\\
	&=\exp(-\theta_k{d_{e,k}^\alpha}),
	\end{split}
	\end{equation}
where 
\begin{equation}
\theta_k \triangleq \frac{1+\tau_k{p_k}+\sum\nolimits_{l\ne 
k}{\beta_{k,l}}{\tau_l}{p_l}-\left(1+\sum\nolimits_{l\ne 
k}{\beta_{k,l}}{\tau_l}{p_l}\right){\delta_{s,k}}}{\left(1+\sum\nolimits_{l\ne 
k}{\beta_{k,l}}{\tau_l}{p_l}\right)\delta_{s,k}p_k}{\sigma_e^2},
\end{equation}
with $\delta_{s,k} \triangleq 2^{R_{s,k}} $. The last equality in (12) follows 
by realizing that $ \left|h_{e,k}\right|^2 $ is exponentially distributed with 
parameter $ d_{e,k}^\alpha $.

With (12), the secrecy constraint (10e) can be rewritten as
\begin{equation}
\label{secrecytran}
\mathrm{exp}\left(-\theta_k{d_{e,k}^\alpha}\right)\le \epsilon.
\end{equation}

Similarly as done for constraint (\ref{chca}), we introduce auxiliary variables $ \left\{\phi_k,u_k,w_k\right\} $ such that (\ref{secrecytran}) can be equivalently transformed as follows

\begin{subequations}
\label{fiveineq}	
\begin{align}
\exp(-\phi_k{d_{e,k}^\alpha})&\le \epsilon,\\
\pi_k\delta_{s,k}&\le u_k,\\
\phi_k{w_k}&\le {(\pi_k+\tau_kp_k-u_k)},\\
p_k{u_k}&\le w_k,\\
1+\sum\limits_{l\ne k}{{{\beta }_{k,l}}{{\tau }_{l}}{{p}_{l}}}&\le {{\pi }_{k}}.
\end{align}
\end{subequations}

Next, we deal with constraint (\ref{order}), which can be rewritten in the following form

\begin{align}
\label{binary}
\beta_{k,l}&\in \left\{0,1\right\},\\
\beta_{k,l}\tau_lp_l& < \tau_kp_k.
\end{align}

In order to handle the binary variable $ \beta_{k,l} $, we further introduce the auxiliary variables $ \mu_{k,l} $ and obtain an equivalent form for (\ref{binary})
\begin{align}
\beta_{k,l}({1-\mu_{k,l}})&=0,\\
\beta_{k,l}&=\mu_{k,l}.
\end{align}

We can easily confirm that the above equality constraints hold when $ \beta_{k,l}\in \left\{0,1\right\} $, which means there is no relaxation from (\ref{binary}) to (18) and (19). In other words, the actual feasible region of the solution to problem (\ref{optimal problem}) would not change if we replace (\ref{binary}) with the above two equality constraints, which are much easier to tackle in the PDD algorithm as will be detailed in the next subsection.

Following the above steps, we can reformulate problem (\ref{optimal problem}) as follows
\begin{subequations}
	\label{optimal_convert}
\begin{align}
\min_\mathcal{V} \quad  &\sum\limits_{k}{{{\varsigma 
		}_{k}}{{c}_{k}}^{3}l_{k}^{3}/{{T}^{2}}+{{p}_{k}}T},\\
\mathrm{s.t.} \quad
&(10\mathrm{b}), (10\mathrm{d}), (10\mathrm{f}), (10\mathrm{g}), (10\mathrm{i}),\\
& (11\mathrm{a})-(11\mathrm{c}),(15\mathrm{a})-(15\mathrm{e}), (17)-(19), 
\end{align}
\end{subequations}
where $ 
\mathcal{V}=\left\{\boldsymbol{l},\boldsymbol{R_t},\boldsymbol{R_s},\boldsymbol{p},\boldsymbol{b},\bm{\beta},\bm{\mu},
 \boldsymbol{\pi},\boldsymbol{\phi},\boldsymbol{u},\boldsymbol{w}\right\} $. 
Note that the 
object function in problem (20) is a continuously differentiable function, the variable set $ \mathcal{V} $ is  
closed convex, the functions in the inequality constraints, e.g., (10b), (10d) and (10f), are all 
differentiable, and the equality constraints, e.g., (18), (19), are continuously 
differentiable. These motivate us to employ the PDD method to solve the above problem \cite{ref12}.
\subsection{Algorithm}
In this subsection, we adopt the SCA and PDD methods to tackle problem (\ref{optimal_convert}). The PDD method utilizes a double-loop structure. First, in order to handle the equality constraints, e.g., (10i), (18), (19), we incorporate the corresponding augmented Lagrangian (AL) terms into the objection function. Then, in the inner loop, we solve the AL problem by applying the SCA method with fixed penalty parameter and dual variable. In the outer loop, we update the penalty parameter or the dual variable according to the constraint violation. The optimization procedure is detailed as follows. 
\subsubsection{AL problem}
We deal with the aforementioned equality constraints by integrating the corresponding AL terms into (20a) which yields the following AL problem
\begin{subequations}
\label{optimal3}
\begin{align}
\begin{split}
\underset{\mathcal{V}}{\mathop{\text{min}}}\,\text{  }&\sum\limits_{k} \left({{\varsigma 
	}_{k}}{{c}_{k}}^{3}l_{k}^{3}/{{T}^{2}}+{{p}_{k}}T\right) \\ &\text{ 
+}\frac{\text{1}}{\text{2}\rho 
}\sum\limits_{k=1}^{K}{\sum\limits_{l=1}^{K}{\left( {{\left| {{\beta 
}_{k,l}}-{{\mu }_{k,l}}+\rho {{\lambda }_{1,k,l}} \right|}^{2}} \right)}}\\ 
&+\frac{\text{1}}{\text{2}\rho 
}\sum\limits_{k=1}^{K}{\sum\limits_{l=1}^{K}{\left( {{\left| {{\beta 
}_{k,l}}(1-{{\mu }_{k,l}})+\rho {{\lambda }_{2,k,l}} \right|}^{2}} 
\right)}}\\&+\frac{\text{1}}{\text{2}\rho 
}\sum\limits_{k=1}^{K}{\sum\limits_{l=1}^{l<k}{\left( {{\left| {{\beta 
}_{k,l}}+{{\beta }_{l,k}}+\rho {{\lambda }_{3,k,l}} \right|}^{2}} \right)}},\\
\end{split}\\
\mathrm{s.t.} \quad
&(10\mathrm{b}), (10\mathrm{d}), (10\mathrm{f}), (10\mathrm{g}),\\  &(11\mathrm{a})-(11\mathrm{c}),(15\mathrm{a})-(15\mathrm{e}), (16),\\
&0 \le \beta_{k,l} \le 1,
\end{align}
\end{subequations}
where $ \rho $ and $ \mathbf{\lambda} $ denote the scalar penalty parameter and 
dual variable, respectively. In addition, constraint (21d) has no influence on 
the optimality and is introduced to improve the convergence speed. In each 
outer loop, the dual variable $ \mathbf{\lambda} $ is updated when the equality 
constraint violation is below a certain level, and otherwise the penalty 
parameter $ \rho $ is updated. When $ \rho \rightarrow 0 $, solving problem 
(\ref{optimal3}) yields an identical solution to problem 
(\ref{optimal_convert}).
\subsubsection{Solving problem ($ \mathrm{21} $)}
In the inner loop, we aim to solve the AL problem (21) by fixing the values of $ \rho $ and $ \mathbf{\lambda} $. Although 
the fractional form, binary variables, and equality constraints have been avoided in problem (21), there are 
still several non-convex constraints, such as (11b), (11c), etc, which makes 
the problem complicated to address. To this end, we employ the SCA method to 
approximate the aforementioned non-convex constraints as convex ones. Take 
constraint (11c) as an example, and we note that the function $ b_k{\pi_k} $ in 
(11c) is jointly concave with respect to $ b_k $ and $ \pi_k $. Accordingly, by 
applying the first-order Taylor expansion around $ (b_k^i,\pi_k^i) $, a convex 
upper bound approximation for $ b_k{\pi_k} $ can be obtained as below
\begin{equation}
b_k{\pi_k} \le {b_k^i}{\pi_k^i} + {\pi_k^i}(b_k-b_k^i)+{b_k^i} (\pi_k-\pi_k^i),
\end{equation}
where $ b_k^i $ and $ \pi_k^i $ denote the values of $ b_k $ 
and $ \pi_k $ in the $ i $-th inner loop. As a consequence, constraint (11c) can be 
approximated by a more stringent but convex constraint given below
\begin{equation}
\label{a}
 {b_k^i}{\pi_k^i} + {\pi_k^i}(b_k-b_k^i)+{b_k^i} (\pi_k-\pi_k^i)\le 
 {\tau_{k}}{p_k}.
\end{equation}

Quite similarly, we can further rewrite the constraints (11b), 
(15b)-(15d), and (17) as below
\begin{align}
\label{b}
&\sum\limits_{l\ne k}{\tau_l}\left[ \beta _{k,l}^{i}\text{ 
}p_{k}^{i}+\beta _{k,l}^{i}({{p}_{k}}-p_{k}^{i})+{{p}_{k}}^{i}({{\beta 
}_{k,l}}-\beta _{k,l}^{i}) \right]\le {{\pi }_{k}}-1,\\
\label{c}
&\pi _{k}^{i}\delta_{s,k}^{i}+\pi _{k}^{i}\left( 
{{\delta}_{s,k}}-\delta_{s,k}^{i} 
\right)+\delta_{s,k}^{i}({{\pi }_{k}}-\pi _{k}^{i})\le {{u}_{k}},\\
\label{d}
&\phi _{k}^{i}w_{k}^{i}+\phi _{k}^{i}\left( {{w}_{k}}-w_{k}^{i} 
\right)+w_{k}^{i}({{\phi }_{k}}-\phi _{k}^{i})\le {{\pi }_{k}}+{{\tau 
}_{k}}{{p}_{k}}-{{u}_{k}}, \\
\label{e}
&p_{k}^{i}u_{k}^{i}+p_{k}^{i}\left( {{u}_{k}}-u_{k}^{i} 
\right)+u_{k}^{i}({{p}_{k}}-p_{k}^{i})\le {{w}_{k}},\\
\label{f}
&p_{l}^{i}\beta_{k,l}^{i}+p_{l}^{i}\left( {{\beta_{k,l}}}-\beta_{k,l}^{i} 
\right)+\beta_{k,l}^{i}({{p}_{l}}-p_{l}^{i})\le p_k\tau_{k}/\tau_{l}.
\end{align}

Finally, the AL problem can be presented as the following convex form
\begin{subequations}
\label{p3}
\begin{align}
\min_{\mathcal{V}}\qquad &(21\mathrm{a}),\\
\mathrm{s.t.} \quad
&(10\mathrm{b}), (10\mathrm{d}), (10\mathrm{f}), (10\mathrm{g}),\\& (11\mathrm{a}), (15\mathrm{a}), (21\mathrm{d}), (\ref{a})-(\ref{f}).
\end{align}
\end{subequations}

In order to solve the above AL problem, we divide the 
variable set $ \mathcal{V} $ into two blocks and update them alternatively. The first 
block contains variable $ \mu_{k,l} $ which only appears in the objective function. Thus, the closed-form solution of the optimal $ \mu_{k,l} $ can be obtained easily. The remaining variables $ \tilde{\mathcal{V}} \triangleq \mathcal{V}\backslash\mu_{k,l}$ are relegated to the second block, and we resort to the convex programming toolbox CVX to calculate the solution. Therefore, the solution for the $ i $-th inner loop can be decomposed as two steps:

Step 1: By fixing the variables $ \tilde{\mathcal{V}} $ in the second block, we can derive the closed-form solution of  $ \mu_{k,l} $ expressed as
\begin{equation}
{{\mu }_{k,l}}=\frac{{{\beta }_{k,l}}+\beta _{k,l}^{2}+\rho {{\lambda 
}_{1,k,l}}+\rho {{\lambda }_{2,k,l}}{{\beta }_{k,l}}}{1+\beta _{k,l}^{2}}.
\end{equation}
 
Step 2: In order to update the variables $ \tilde{\mathcal{V}} $, we fix $ \mu_{k,l} $ and 
utilize CVX to solve the AL problem (29).
\subsubsection{The overall algorithm}
The overall algorithm is summarized in Algorithm 1, where $ 
{f}_{\tilde{\mathcal{V}}}^i $ denotes the object function in the $ i $-th inner 
loop, and $ {g}_{\tilde{\mathcal{V}}}^j $ denotes the vector that combines all 
functions in the equality constraints of problem (29) in the $ j $-th outer 
loop. The convergence of the proposed algorithm can be proved similarity as 
done in \cite{ref12}, which is omitted here due to page limitation. 
\begin{algorithm}
	\caption{PDD-based algorithm for solving problem (\ref{optimal problem})}
	\begin{algorithmic}
		\STATE Initialize $ \delta $, $ I_{max} $, $ \rho_0 $, $ \lambda_0 $, $ 
		j=0 $, $ i=0 $ and $ 0<c<1$. Initialize a feasible point for the 
		algorithm.
		\REPEAT
		  \REPEAT
		  \STATE Update $ \mu_{k,l} $ based on (30).
		  \STATE Update $ \tilde{\mathcal{V}} $ by solving problem (29) using CVX toolbox.
		  \STATE Update $ i=i+1 $.
		  \UNTIL $ \frac{\left| {{f}_{\tilde{\mathcal{V}}}^{i+1}}-{{f}_{\tilde{\mathcal{V}}}^i} 
		  \right|}{\left| {{f}_{\tilde{\mathcal{V}}}^i} \right|}\le {{\delta }} 
		  $, or $ i>I_{max} $.
         \IF {$ {{\left\| {{g}_{\tilde{\mathcal{V}}}^j} \right\|}_{\infty }}\le {{\eta 
         }_{j}} $}
           \STATE $ {{\lambda }^{j+1}}={{\lambda }^{j}}+{g}_{\tilde{\mathcal{V}}}^j/\rho^j $
            \STATE$ \rho^{j+1} =  \rho^{j} $
         \ELSE
         \STATE$\lambda^{j+1}=\lambda^{j}$
         \STATE$ \rho^{j+1}=c\rho^{j} $
         \ENDIF
         \STATE Update $ j=j+1 $.
		\UNTIL $ {{\left\| {{g}_{\tilde{\mathcal{V}}}^j} \right\|}_{\infty }}\le \delta$.
	\end{algorithmic}
\end{algorithm}

\underline {\emph{Complexity analysis}}: The complexity of Algorithm 1 is dominated by solving problem (29) which contains $ 6K $ first-order Taylor expansion constraints. The number of variables is  $ (9K+K^2) $. Therefore, the complexity is on the order of $ I_1I_2O(K^3) $, where $ I_1 $ and $ I_2 $ denote the number of the inner and  outer iterations, respectively.

\section{Numerical Results}

In this section, numerical results are provided to evaluate the convergence and 
performance of the proposed algorithm based on SCA and PDD. Unless otherwise 
specified, we set the number of users $ K=3 $, system bandwidth $ 
B=10\mathrm{MHz} $, duration $ T=0.1\sec $, pass-loss exponent $ \alpha=5 $, 
noise variance $\sigma_{b}^2=\sigma_e^2=-50\mathrm{dBm} $, CPU cycles $ 
C_k=10^3 $ cycles/bit, effective capacitance coefficient $ 
\varsigma_{k}=10^{-28} $, distance $ d_{e,k}=100\mathrm{m} $, secrecy outage 
probability threshold $ \epsilon=0.1 $, and the tolerance error $ \delta = 
10^{-4} $.
\begin{figure} 
	\centering 
	\includegraphics[scale=0.6]{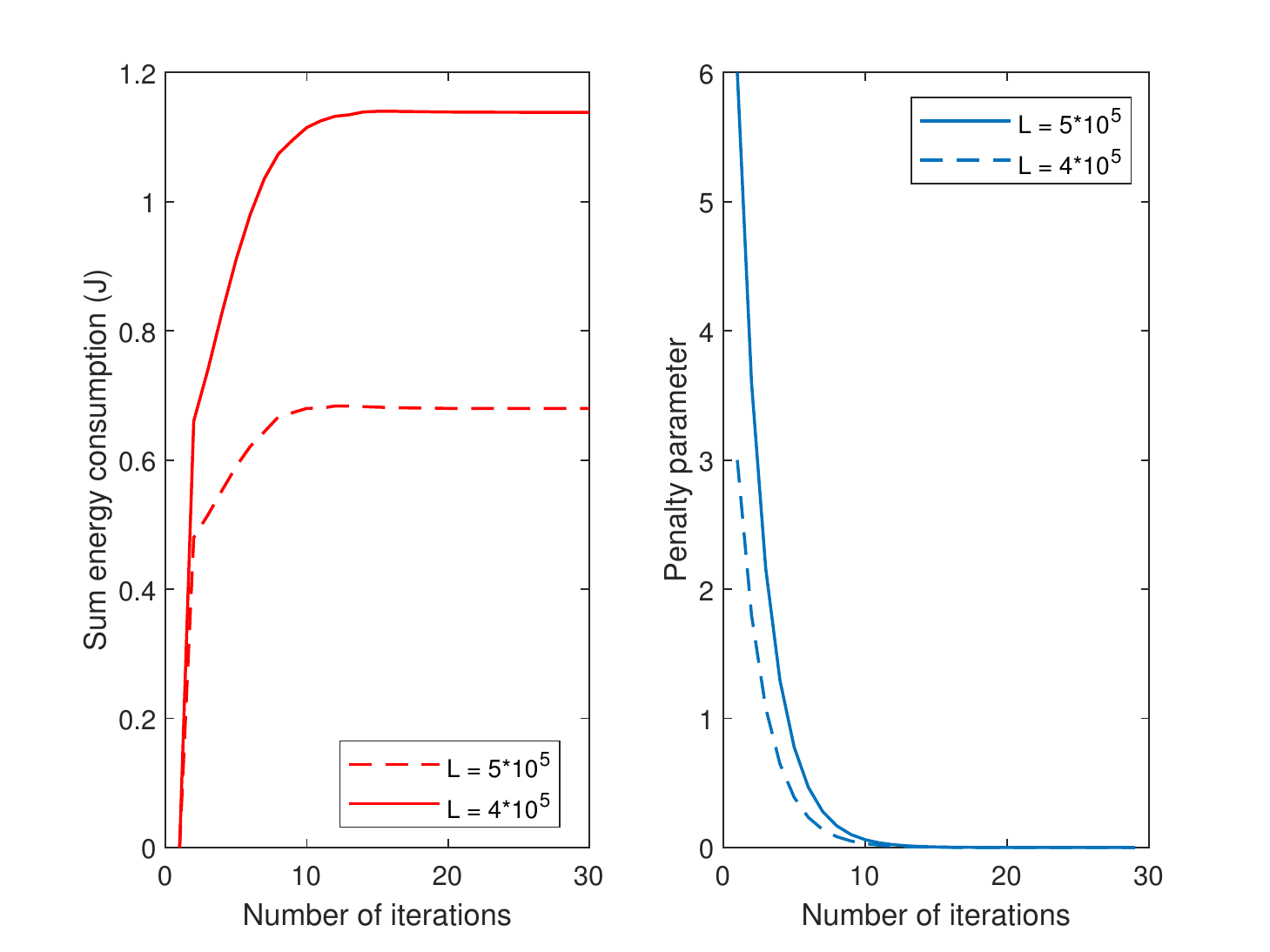} 
	\caption{Sum energy consumption and penalty parameter $ \rho $ vs. the number of iterations.} 
\end{figure} 

\begin{figure}[h]
	\centering
	\includegraphics[scale=0.6]{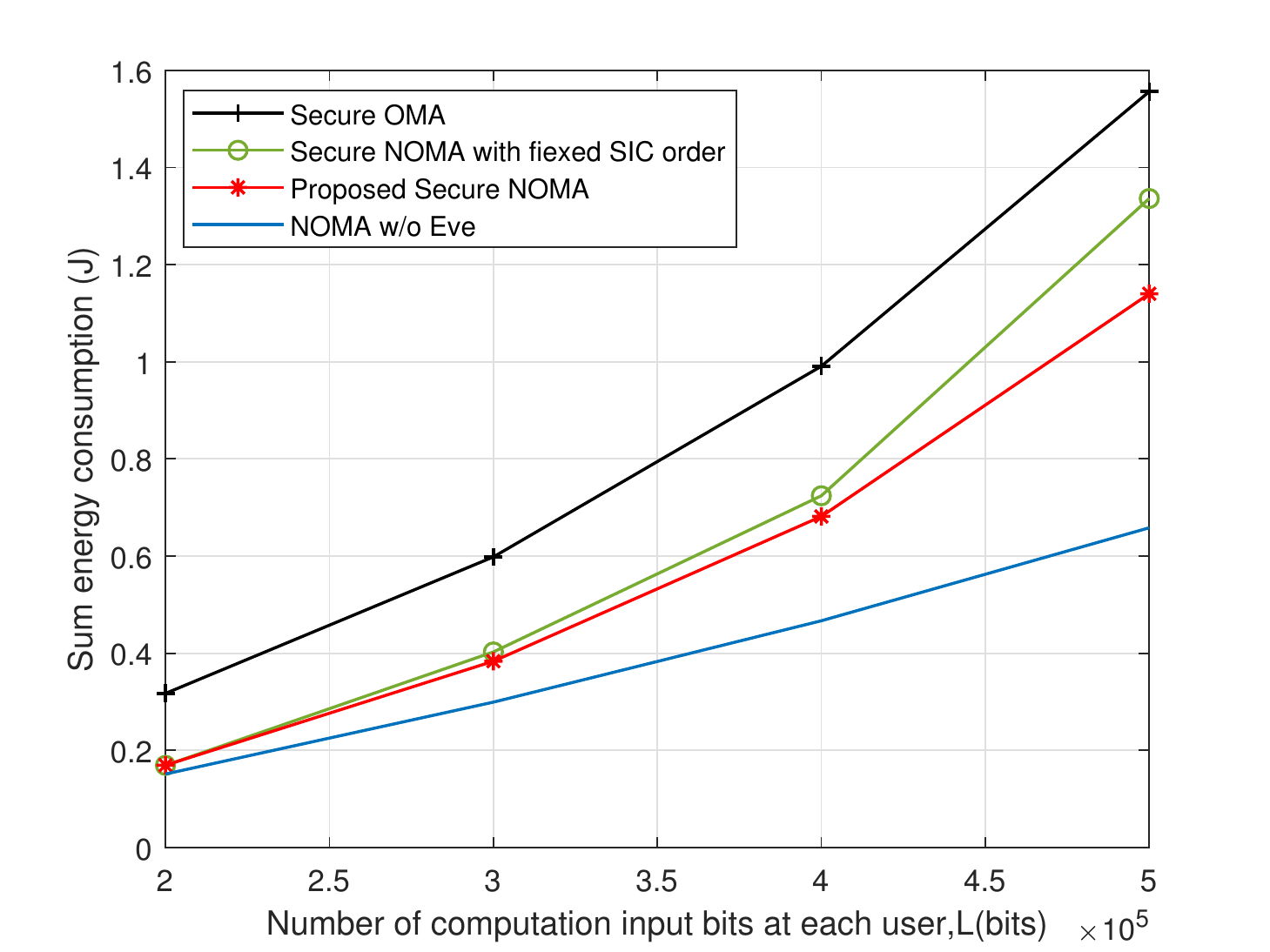}
	\caption{Sum energy consumption vs. the number of computation input bits $ 
	L $, $ K=3 $.}	
\end{figure}
First, we examine the convergence of the proposed algorithm for computation input bits $ L=4\times10^5 $ bits and $ L=5\times10^5 $ bits where we set the initial penalty parameter $ \rho^0=10 $, the decrease number $ c = 0.6 $, and the equality constraint violation tolerance parameter $ \eta_{j} =0.3^j $ in the $ j $-th outer iteration. Fig. 2 shows the sum energy consumption and the value of penalty parameter versus the number of iterations. We observe that the curves of sum energy consumption in the left figure reach saturation quickly which verifies the convergence of the proposed algorithm. We can also find that the penalty parameter shown in the right figure sharply decreases to zero. The above results justify well the effectiveness of our proposed algorithm for addressing the sophisticated problem (\ref{optimal problem}).

Fig. 3 compares the sum energy consumption versus the number of 
computation input bits $ L_k=L $ of our scheme with the following 
three benchmark schemes:

\textit{Secure OMA}: The $ K $ users adopt the time division multiple access 
(TDMA) protocol for computation offloading. For simplicity, we assume that the 
duration $ T $ is equally divided to the users.

\textit{NOMA without Eve}: There does not exist any eavesdropper in the 
network, and therefore we do not need to consider the secrecy constraint for 
computation offloading.

\textit{Secure NOMA with fixed SIC order}: As for the SIC decoding order, we 
simply use the descending order of users' channel gains which can avoid the 
binary variables.

It is observed from Fig. 3 that by introducing the advanced NOMA technology, 
our proposed design outperforms the OMA one. At the same time, we can also see 
our proposed design consumes less energy compares with the design with fixed 
SIC decoding order. This implies that simply using the descending order of 
users' channel gains as the SIC decoding order in the NOMA scheme, as done in 
\cite{ref8}, might produce a highly suboptimal solution. This also manifests 
the significance of designing the decoding order of the NOMA scheme for MEC 
networks. Besides, for the purpose of anti-eavesdropping, the proposed design 
consumes more energy than the one without Eve, which implies that the secrecy 
is achieved at the cost of more energy consumption.
\section{Conclusions}
In this paper, we considered the security issue in an uplink NOMA-MEC system in the presence of a 
malicious Eve, where $ K $ users simultaneously offload partial of their 
computation tasks to the BS using the NOMA technique. In order to solve the 
non-convex optimization problem of minimizing the sum energy consumption, we 
proposed an algorithm based on SCA and PDD and jointly design the optimal  
local computing bits, codeword rate, confidential rate, transmit power, and SIC 
decoding order. Numerical results show that the NOMA schemes significantly 
outperforms the one with OMA. In addition, designing the optimal SIC decoding order can further improve the performance. 
%%\section*{Acknowledgment}                                        
%This work was partially supported by the National Natural Science Foundation of China under Grant 61701390, the Open Research Fund of National Mobile Communications Research Laboratory, Southeast University, under Grant 2021D07, and the Post-Doctoral Research Project of Shaanxi Province under Grant %	2017BSHYDZZ38.

\end{document}